\documentclass[a4paper]{JHEP3} 
\usepackage{microtype}
\usepackage{lmodern}
\usepackage[T1]{fontenc}
\usepackage{color}
\let\normalcolor\relax 

\usepackage{graphicx}
\usepackage{dcolumn}
\usepackage{bm}
\usepackage{amssymb}
\usepackage{mathrsfs}
\usepackage{amsmath}
\usepackage{epsfig}
\usepackage{amsmath}
\usepackage{rotating}
\usepackage{timestamp} 

\def\be{\begin{equation}}
\def\ee{\end{equation}}

\def\bea{\begin{eqnarray}}
\def\eea{\end{eqnarray}}

\title{Constraints on TeV scale Majorana neutrino phenomenology from the Vacuum Stability of the Higgs}

\author{Joydeep Chakrabortty,
Moumita Das,
Subhendra Mohanty
\\
   Physical Research Laboratory, Ahmedabad-380009, India
 \\
\email{joydeep@prl.res.in},
\email{moumita@prl.res.in},
\email{mohanty@prl.res.in}

}

\abstract
{ The vacuum stability condition of the Standard Model Higgs potential with mass in the range of 124-127 GeV puts an upper
bound on the Dirac mass of the neutrinos. We study this constraint
with the right-handed neutrino masses upto TeV scale. The heavy neutrinos contribute to $\Delta L=2$ processes like 
neutrinoless double beta decay and same-sign-dilepton production in the colliders.
The vacuum stability criterion also restricts the light-heavy neutrino
mixing and constrains the branching ratio of lepton flavour violating process, like $\mu \to e \gamma$ mediated by the heavy neutrinos.
We show that neutrinoless double beta decay with a lifetime $\sim 10^{25}$ years can be observed if the the lightest
heavy neutrino mass is $<$ 4.5 TeV. We show that the vacuum stability condition and the experimental bound on 
$\mu \rightarrow e \gamma$ together put a constrain on heavy neutrino mass $M_R >$ 3.3 TeV.  
Finally we show that the observation of same-sign-dileptons (SSD) associated with jets at the LHC
needs much larger luminosity than available at present. 
We have estimated the possible maximum cross-section for this process at the LHC and show that with an integrated luminosity
100 $fb^{-1}$ it may be possible to observe the SSD signals as long as $M_R <$ 400 GeV. }

\begin{document}
 
\section{Introduction}
The recent measurement of ATLAS and CMS \cite{Atlas-cms, ATLAS:2012ae,Chatrchyan:2012tx} have confirmed the existence of a new boson which has mass in the range
 $126.5$ GeV (ATLAS at 5.0$\sigma$)
and $125.3 \pm 0.6$ GeV (CMS at 4.9$\sigma$), and it is expected to be a Standard Model Higgs. 
This mass range implies that quartic coupling $\lambda_h$ of the Higgs has a value close to the vacuum stability
limit \cite{Altarelli:1994rb,Holthausen:2011aa,EliasMiro:2011aa,Xing:2011aa,Chetyrkin:2012rz,Bezrukov:2012sa,Degrassi:2012ry}. The top-quark
loop makes a negative contribution to the $\beta$-function of $\lambda_h$ while the gauge couplings give a
positive contribution. If the quartic coupling $\lambda_h(\mu)$ becomes negative at large renormalization scale $\mu$,
it implies that in the early universe the Higgs potential would be unbounded from below and the vacuum would be unstable
in that era.
It has been pointed out that the Higgs mass in the 126 GeV range being close to the vacuum stability limit, one can put stringent
constraints on new physics which affects the running of the Higgs quartic coupling.

One class model which can be constrained from the stability criterion of the Higgs coupling is the see-saw models of neutrino
masses \cite{Casas:1999cd,Gogoladze:2008gf,Gogoladze:2008ak,Chen:2012fa,Pilaftsis:1991ug,Rodejohann:2012px}. In Type-I see-saw models 
\cite{Type-1} one introduces a number of heavy gauge singlet Majorana neutrinos which have Yukawa couplings with the
Higgs and lepton doublets. The electroweak symmetry breaking gives rise to the Dirac mass matrix $\mathcal{M}_D$,
\be
-\mathcal{L}=\bar{N_{R}}{\mathcal M}_{D}\nu_{L}+\frac{1}{2}\bar{N_{R}}{\mathcal M}_{R}N_{R}^{c}+\rm{h.c.}
\ee
If $\mathcal{M}_D << \mathcal{M}_R$ in the pure Type-I models \cite{Type-1} the light neutrino masses are given by
${\mathcal M}_\nu= {\mathcal M}_D^T {\mathcal M}_R^{-1} {\mathcal M}_D$.
It has been discussed earlier in many papers that light neutrino masses which can explain the solar and atmospheric
neutrino oscillations are obtained by assuming the eigenvalues $M_D \sim 100$ GeV and $M_R > 10^{14}$ GeV. 
By a suitable choice of ${\mathcal M}_D$ and ${\mathcal M}_R$ one can set  ${\mathcal M_D^T \mathcal{M}_R^{-1} \mathcal{M}_D}=0$ and the 
light neutrino masses are given by higher order terms in  ${\mathcal M_D^T \mathcal{M}_R^{-1}}$
\cite{Grimus:2000vj,Kersten:2007vk,Hettmansperger:2011bt}. 
In this way it is possible to generate viable light neutrino masses while reducing the scale $M_R$ to less than a TeV. 
 In \cite{Pilaftsis:1991ug,Rodejohann:2012px}, the constraints on various TeV-scale
Type-I neutrino mass models from the vacuum stability criterion of Higgs coupling has been checked.

In this paper we assume a Yukawa couplings of heavy Majorana neutrinos with the lepton doublets with the Standard Model
Higgs and that the heavy neutrinos masses are in the 100 GeV-10 TeV range. We obtain the constraints on the
Higgs-neutrino Yukawa couplings by calculating the renormalisation group evolutions (RGEs) of $\lambda_h(\mu)$ (which is fixed at the
electroweak scale by the Higgs mass). The vacuum stability condition is the requirement that
$\lambda_h(M_W \leq \mu \leq M_P) \geq 0$. We find that this leads to the constraint $Y_\nu \leq 0.14$ on the elements
of the Yukawa coupling matrix. We then apply this condition (which implies that the  Dirac neutrino masses
$M_D \leq 24.36$ GeV) on the phenomenology of TeV scale heavy neutrinos \cite{Atre:2009rg,Ibarra:2010xw}.

We study three aspects of the heavy neutrino phenomenology in the light of the vacuum stability condition on $Y_\nu$:
(1) Neutrino-less double beta decay ($0 \nu \beta \beta$), (2) Lepton flavor violating decays like
$\mu \rightarrow e \gamma$, and (3) Same-sign-dilepton signals at the LHC. 
All these process depend upon the mixing of the light neutrino gauge eigenstates with the 
heavy neutrino mass eigenstates which is given by the mixing matrix $V \simeq {\mathcal M_D^T}{\mathcal M_R^{-1}}$. 
The Dirac mass ${\mathcal M_D}$ gets an upper bound from the vacuum stability criterion which in turn puts constraints on the processes listed above. 
Our analysis is independent of the specific light neutrino mass model. 
We shall put upper bounds on various processes by assuming that  the  elements of ${\mathcal M_D}$ which contribute 
to that particular signal are as large as can be allowed by the vacuum stability condition. Our choices 
for ${\mathcal M_D}$ and ${\mathcal M_R}$ may not give realistic neutrino mass through the usual Type-I seesaw which means that these signals 
can be further restricted by specific choice of the flavour structures.

In Section (2) we discuss the running of the Higgs quartic coupling in the Standard Model for the 125 GeV Higgs.
In Section (3) we introduce the Yukawa couplings $Y_\nu$ between the Higgs and heavy neutrinos in the context of
SM extended by heavy singlet fermion and study the effect of these neutrino Yukawa's on the running of the Higgs quartic coupling.
In this section we establish the bound on $Y_\nu$ from the stability criterion. In Section (4) we study
$(0 \nu \beta \beta)$, in Section (5) the lepton flavor violations, and in Section (6) we estimate the same-sign-dilepton 
signals at the LHC. Our results are summarized in the concluding section.

\section{ Vacuum stability of the Standard Model Higgs potential}
The Higgs mass measured by ATLAS and CMS collaborations \cite{Atlas-cms, ATLAS:2012ae,Chatrchyan:2012tx}is in the mass range 124.7 GeV-126.5 GeV  
is close  to the bound on Higgs mass from of electro-weak vacuum stability condition \cite{Altarelli:1994rb}. In \cite{Holthausen:2011aa},
it has been shown that in Standard Model, the Higgs boson quartic coupling $\lambda_h$ can remain positive
upto Planck scale with appropriate choice of top quark mass $m_t$, strong coupling constant $\alpha_s$ etc.
The coupling $\lambda_h > 0$ ensures the stable vacuum and the bounded potential from below. Details of
this study is performed recently in \cite{Degrassi:2012ry}.

The RG-improved Higgs quartic term can be written as,
\begin{eqnarray}
V_{\rm{eff}}=\frac{\lambda_h (t)}{4!}\left[\xi(t)\phi\right]^4,
\label{effective}
\end{eqnarray}
where $\xi$ signifies the wave function renormalisation and $t\sim \log(\mu/M_Z)$, $\mu$ is the scale of
renormalisation. Here $\lambda_h(t)$ is the effective Higgs quartic coupling with loop corrections. Loop corrections can cause an instability of the potential if
$\lambda(t)$ becomes negative at any scale $\mu <M_P$.
The gauge boson loop makes a positive contribution whereas the top quark makes a negative contribution to the 
$\beta$ function of $\lambda_h$. Hence the instability of the potential mainly comes from loop-correction of top quark.

\begin{figure}[ht]
\begin{center}
\includegraphics[width=0.9\textwidth]{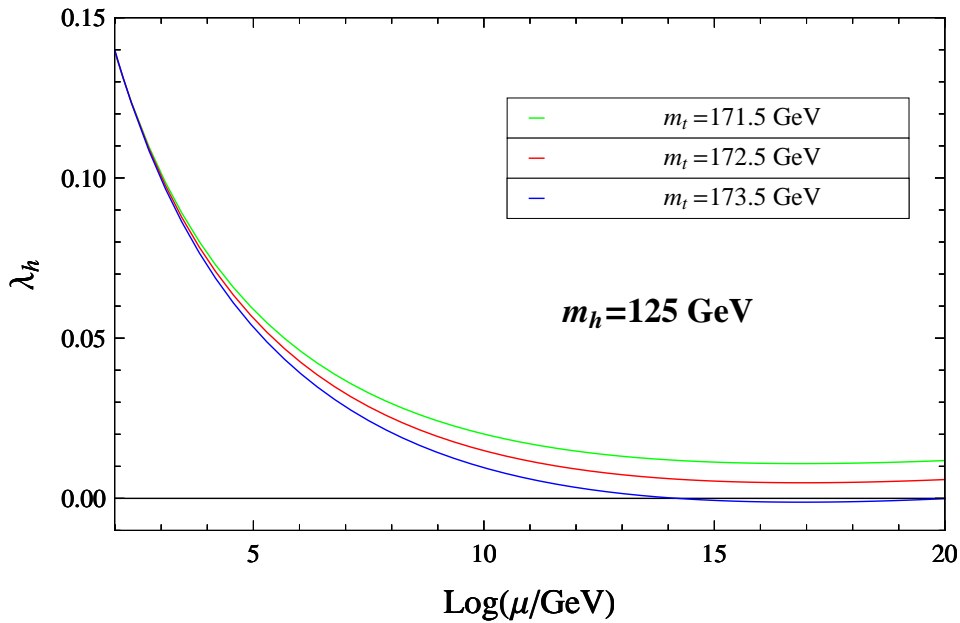}
\end{center}
\caption{\textit{RGE of $\lambda_h$ for different values of $m_t$ in the Standard Model (\rm{with} $m_h=125$ \rm{GeV}, $\alpha_s = 0.1184)$.}}
\label{diff_mtop}
\end{figure}

We compute the RG running of $\lambda_h(\mu)$ using the two loop RG equations available for the Standard Model  \cite{Holthausen:2011aa,Bezrukov:2012sa,Jones}.
We have also inluded the proper matching conditions at top pole mass \cite{Hambye:1996wb}.
The Fig.~\ref{diff_mtop} shows the variation of $\lambda_h$ with different values top quark
mass $m_t$. With increase of $m_t$, $\lambda_h$ becomes negative even before the Planck scale.
For subsequent calculations, we have chosen different sets of top mass keeping Higgs mass constant, and $vice-versa$.

Now we move beyond the Standard Model by adding a heavy neutrino Yukawa coupling.

\begin{figure}[ht]
\begin{center}
\includegraphics[width=0.9\textwidth]{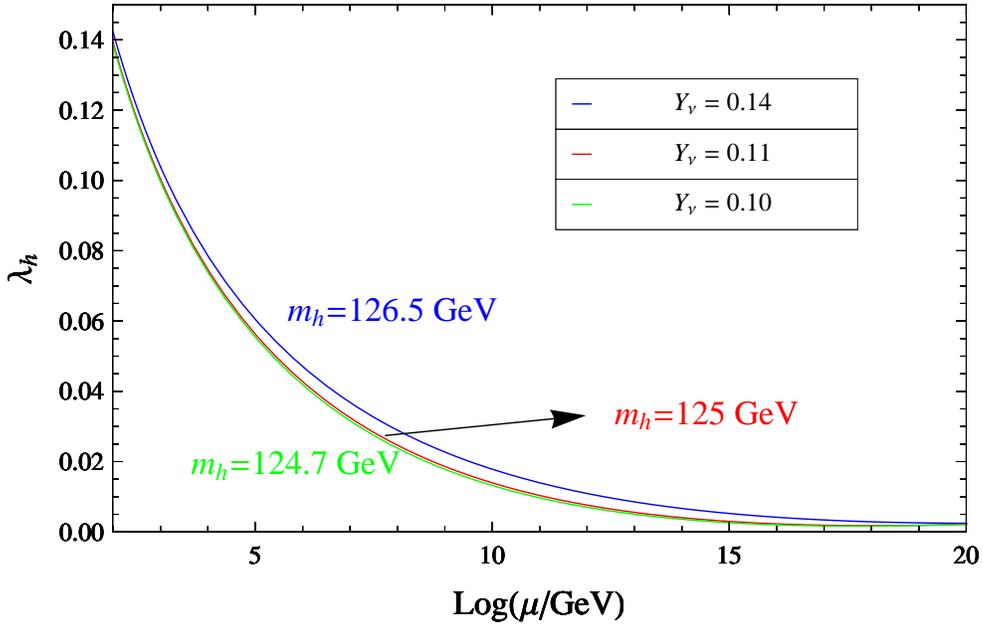}
\end{center}
\caption{\textit{RGE of $\lambda_h$ for different values of $m_h$ in Standard Model (\rm{with} $m_t=172.5$ \rm{GeV}, $\alpha_s = 0.1184)$.}}
\label{diff_mhiggs}
\end{figure}

\section{ Higgs coupling with heavy neutrino}

The Standard Model Higgs can couple to a singlet neutrino $N_R$ via the gauge invariant interaction term
\begin{equation}
-{\cal L}_{Y}= {\mathcal Y_{\nu}} L H N_R +\frac{1}{2}\bar{N_{R}}{\mathcal M_{R}} N_{R}^{c} + \rm{h.c}.,
\end{equation}
where $L=(\nu, l)^T$ is the lepton doublet, $H=(h^0,h^-)$
{\footnote {Once this neutral field $h^0$ acquires vacuum expectation value (vev) v=174 GeV the electroweak symmetry breaking occurs.}}
is the Higgs doublet and $N_R$ is right-handed singlet neutrino.
$\mathcal{M}_R$ are the Majorana masses for $N_R$.
This interaction generates Dirac mass term, $\mathcal{M}_D$, after electroweak symmetry breaking
which reads as $\mathcal{M}_D=\mathcal{Y}_\nu v$ ($v=174$ GeV).

In our further analysis we will not consider the flavour structures of both $\mathcal{Y}_\nu$ and $\mathcal{M}_R$, i.e.,
we will assume that $\mathcal{Y_\nu}=Y_\nu \mbox{diag}(1,1,1)$ and right-handed neutrinos are degenerate, i.e. $\mathcal{M}_R=M_R\,\mbox{diag}(1,1,1)$.

We will see that this new Yukawa coupling affects the RG evolutions of $\lambda_h$ and thus gets constrained from vacuum stability.
This $Y_{\nu}$ also plays important role in the production and decays of $N_R$ leading to same-sign-dilepton associated with jets at the LHC.

The running of neutrino Yukawa coupling is as follows,
\cite{Casas:1999cd,Gogoladze:2008ak,Chen:2012fa}
\begin{eqnarray}
\mu\frac{d}{d\mu}\left(\mathcal{Y}^\dagger_\nu \mathcal{Y}_\nu\right)=\frac{1}{(4\pi)^2}\mathcal{Y}^\dagger_\nu \mathcal{Y}_\nu\left[6 \lambda_t^2 + 2\textrm{ Tr}
\left(\mathcal{Y}^\dagger_\nu \mathcal{Y}_\nu\right) - \left( \frac{9}{10}g^2_1 + \frac{9}{10}g^2_2 \right)+3 \mathcal{Y}^\dagger_\nu \mathcal{Y}_\nu \right]. 
\end{eqnarray}

The introduction of
neutrino sector to Standard Model also modify the RG evolution of the Higgs quartic coupling
$\lambda_h$ and Yukawa coupling of top quark $\lambda_t$ as follows,

 The extra contribution for the singlet fermionic field to Higgs quartic coupling ($\lambda_h$) is
\begin{eqnarray}
\hat{\beta}_{\lambda_h}&=&\frac{1}{(4\pi)^2}\left[-4\textrm{ Tr}(\mathcal{Y}_{\nu}\mathcal{Y}_{\nu}^{\dagger}
 \mathcal{Y}_{\nu}\mathcal{Y}_{\nu}^{\dagger}) +4\lambda_h \textrm{ Tr}(\mathcal{Y}_{\nu}\mathcal{Y}_{\nu}^{\dagger})\right], 
\end{eqnarray}
and to the top quark Yukawa coupling ($\lambda_t$) is
\begin{eqnarray}
\hat{\beta}_{\lambda_t}&=&\frac{1}{(4\pi)^2}\left[\textrm{ Tr}\left(\mathcal{Y}^\dagger_\nu \mathcal{Y}_\nu\right) \right].  
\end{eqnarray}

\begin{figure}[ht]
\begin{center}
\includegraphics[width=0.9\textwidth]{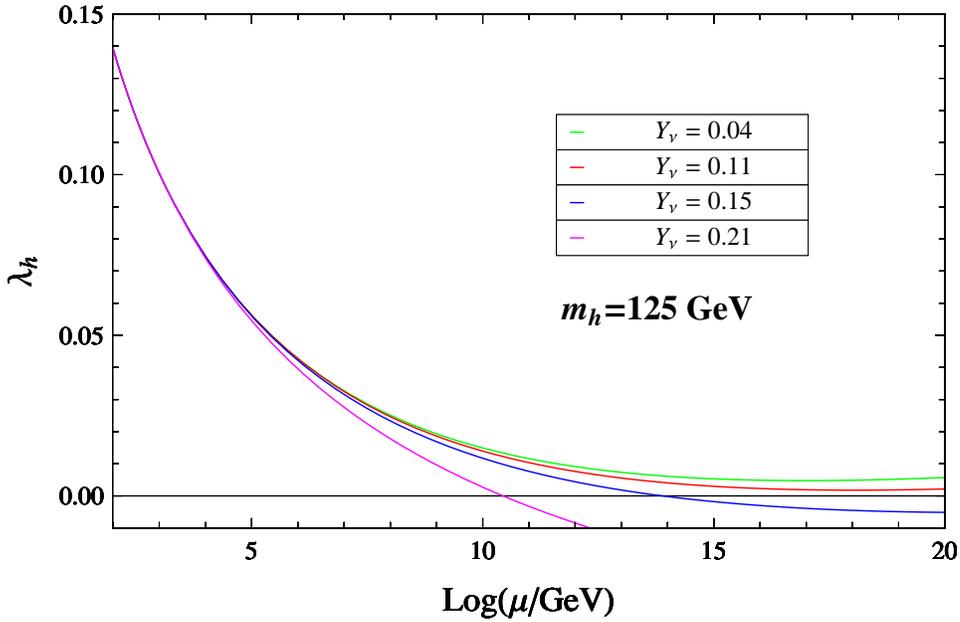}
\end{center}
\caption{\textit{Running of $\lambda_h$ for different values of neutrino Yukawa coupling $Y_{\nu}$
with $M_R = $ 0.1-1 TeV, $(m_t=172.5~ \rm{GeV}, \alpha_s = 0.1184)$.}}
\label{diff_ynu}
\end{figure}

Using these RG equations,
the running of $\lambda_h$ for this model has been shown in Fig.~\ref{diff_ynu}.
The impact of neutrino Yukawa coupling $Y_{\nu}$ on $\lambda_h$ is in similar fashion as the $\lambda_t$ and
$\lambda_h$ becomes negative before Planck scale ($M_{Planck}$) with comparatively larger values of $Y_{\nu}$.
We know there is an uncertainty in top mass measure measurement $173.2\pm 0.9$ GeV \cite{topmass-measurement} and
$173.3\pm 2.8$ GeV \cite{topmass-higgsmass}, and that feature has been grabbed in Fig.~\ref{diff_mtop}.
In Fig.~\ref{diff_mhiggs} we perform
the evolutions of $\lambda_h$ for different Higgs masses choosing suitable top mass.

We outline the RGEs of $\lambda_h$ for different sets of $Y_\nu$ for $m_h= 124.7-126.5$ GeV and $m_t=172.5$ GeV.
We check the stability condition, defined as $\lambda(\mu \leq M_{Planck}) >0$ and reveal that to avoid the
instability of potential, the maximum value of the Yukawa coupling $Y_{\nu}$ at $\mu \sim\rm{TeV}$ must be:
\bea
Y_{\nu} &\le& 0.14.
\label{Ybound}
\eea
This upper limit of $Y_{\nu}$ sets the tolerance of the vacuum in this model.
It has been noted that the light-heavy mixing parameter can be
encapsulated in terms of the Dirac mass, $M_D\sim Y_{\nu}v$, see \cite{Ibarra:2010xw,Mitra:2011qr}.
In other words this mixing which in turn also affects the production and decay of the heavy Majorana neutrino gets constrained.
Thus eventually this bound can be useful to adjudge the possibility of being probed or ruled out this TeV scale model at the LHC.

\section{Gauge interactions of heavy neutrinos }
We consider three generations of Standard Model $SU(2)_L$ lepton doublets $L_{l L}=(\nu_l, \ell_l)_L^T$, $(\ell=e,\mu,\tau)$
and three singlets $N_{R}$.
The relation between the neutrino flavour and the mass eigenstates can be written as
\begin{eqnarray}
\nu_{l L}&=& \sum_{i=1}^3 U_{l i}\nu_{iL} + \sum_{k=1}^{3} V_{l k} N_{k L}^c\\
U^\dagger U &+& V^\dagger V= I,
\end{eqnarray}
where the mixing between the light and heavy neutrinos is $V^\dagger V \simeq (\mathcal{M}_D \mathcal{M}_R^{-1})^2=(v \mathcal{Y}_\nu \mathcal{M}_R^{-1})^2$.
In terms of the mass eigenstates the charged current interaction vertices can be written as
\begin{eqnarray}
-{\cal L}_{int}^{cc} &=& \frac{g}{\sqrt 2} W_\mu\left( \sum_{i=1}^3 U_{l i}^* {\bar \nu_i} \gamma^\mu P_L l
+\sum_{k=1}^{3} V_{l k}^* \overline{N_k^c} \gamma^\mu P_L l  + \rm{h.c.} \right) \nonumber\\
\end{eqnarray}

Our phenomenological studies will involve $\Delta L=2$ processes, like
same-sign-dilepton (including $(0\nu \beta \beta)$) production at colliders where the source of the lepton number violation is the
exchange of heavy Majorana neutrino. The coupling of the heavy neutrino to the charged leptons is parametrised by the mixing angles of $V_{l k}$.
We use the upper bound of $Y_{\nu}$ from Eq. (\ref{Ybound}) to predict the parameter space where these processes may be observable.
We also study lepton flavour violations like $\mu \rightarrow e \gamma$ whose upper limits are again restricted by Eq. (\ref{Ybound}).

\subsection{Neutrinoless double beta decay}

\begin{figure}[ht]
\begin{center}
\includegraphics[width=0.35\textwidth]{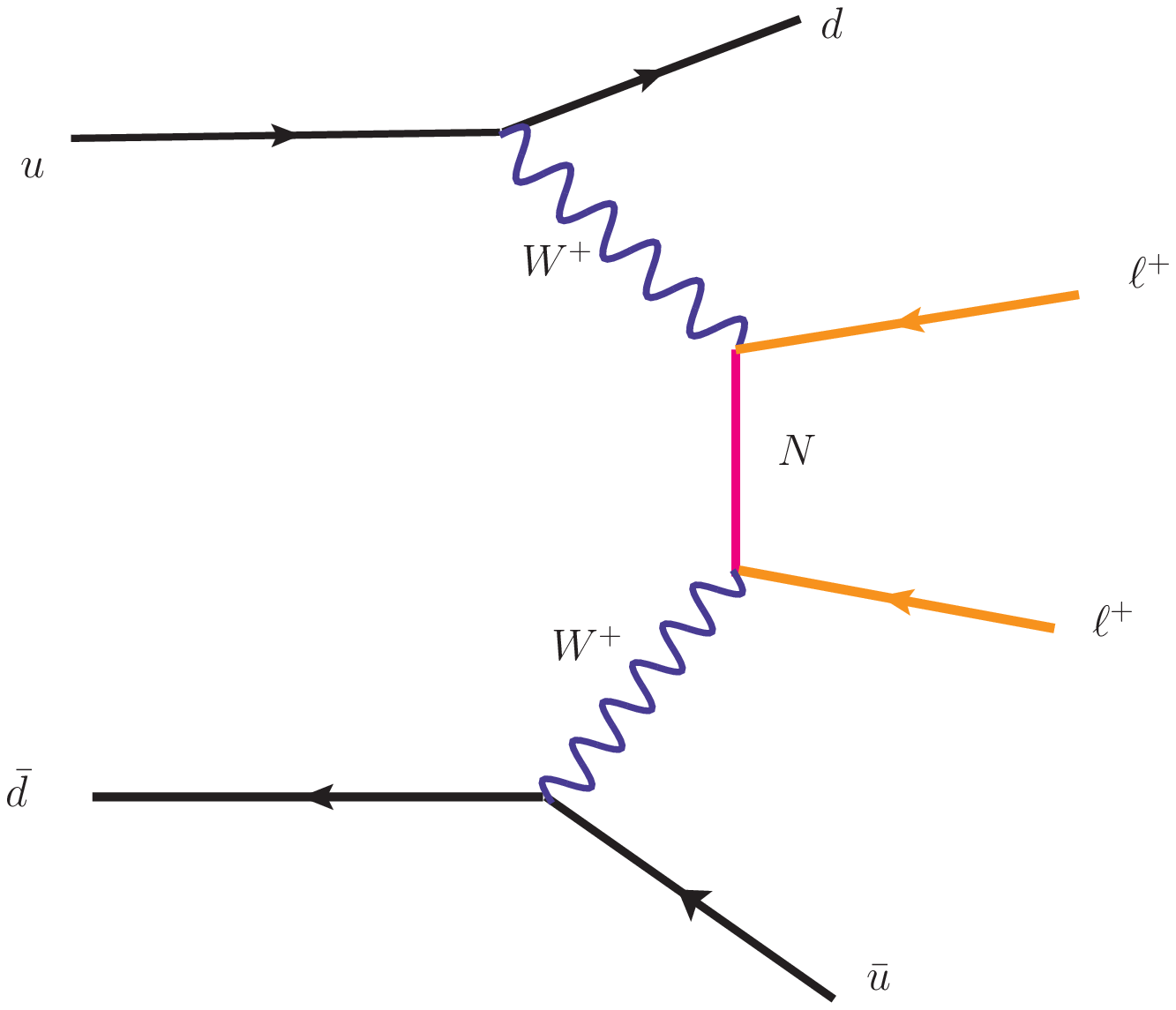}
\hskip 0.01cm
\includegraphics[width=0.55\textwidth]{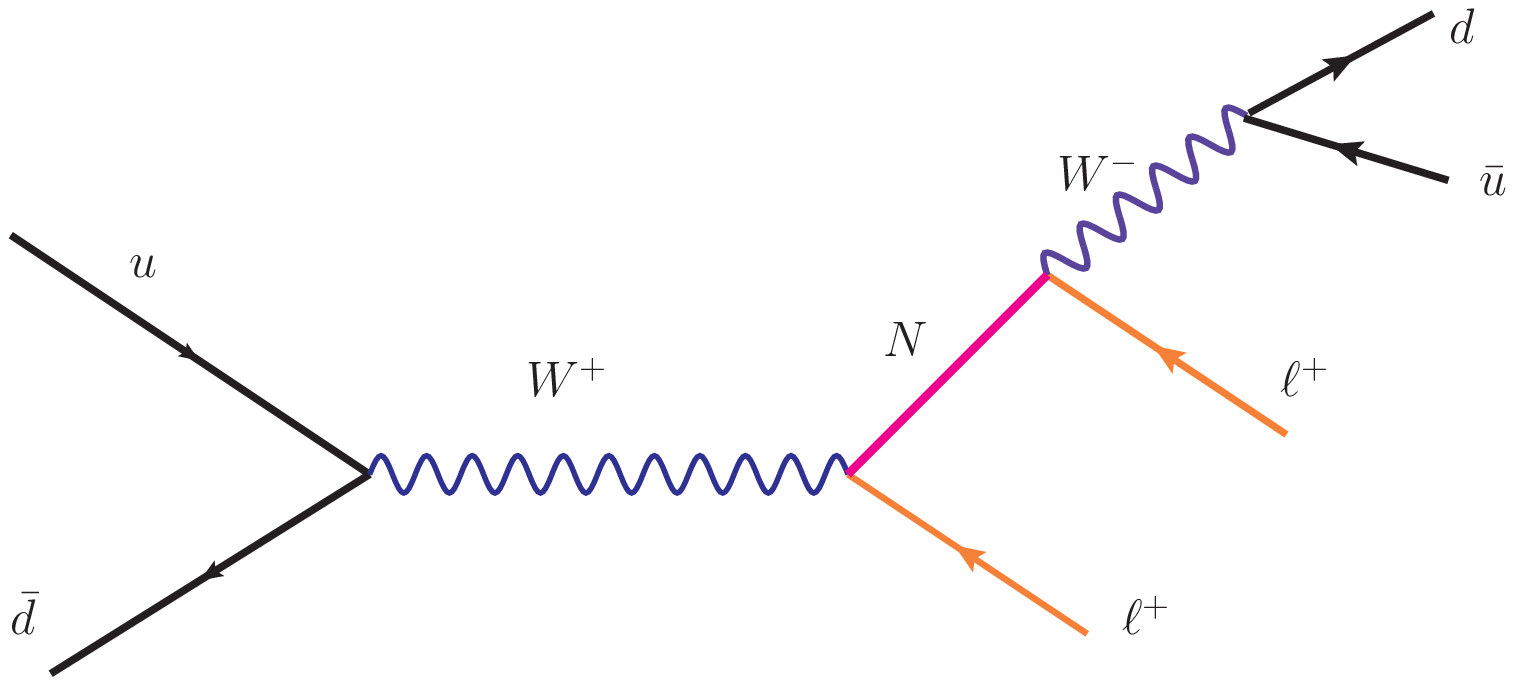}
\end{center}
\caption{\textit{Neutrinoless double beta decay diagrams involving heavy Majorana field.}}
\label{0n2b}
\end{figure}

Neutrinoless double beta decay is one of the important phenomena to probe the lepton number
violation. In this process, the lepton number violation occurs by two units. The half-life time of this process is also
be ascribed by this mixing $V_{li}$ as following:
\begin{eqnarray}
T_{1/2}^{-1}=\kappa_{0\nu}\left|\frac{\left(M_{\nu} \right )_{ee}}{\left\langle p^2\right\rangle}-\frac{|V_{ei}|^2}{M_R }\right|^2,
\label{T_half}
\end{eqnarray}
where $\kappa_{0\nu}=\mathcal{G}_{0\nu}\left(\mathcal{M}_N m_p\right)^2$, nuclear matrix element (NME) for heavy neutrino,
$\mathcal{M}_N =363\pm 44$, $ m_p$ is the proton mass, and $\mathcal{G}_{0\nu}=7.93 \times 10^{-15}~ \rm{yr}^{-1}$.
We assume that the second term arising from the heavy neutrino mixing dominates and
the mixing parameter $V_{ei}$ is explicitly related to the neutrino Yukawa coupling $Y_{\nu}$ via Dirac mass as:
\begin{eqnarray}
\left|V_{ei}\right|^2= \left|\left(M_D M_R^{-1}  \right)_{ee}\right|^2,
\label{Vmix}
\end{eqnarray}
and the relation Eq. (\ref{T_half}) for half-life time for
neutrinoless double beta decay as,
\begin{eqnarray}
T_{1/2}^{-1}  \approx \frac{\kappa_{0\nu}\left|V_{ei}\right|^4}{M_R^2}
=\frac{K_{0\nu}}{M_R^2}\left|\left(M_D M_R^{-1} \right)_{ee}\right|^4.
\end{eqnarray}

The experimental bound on half-life time is $T_{1/2}= 2.23^{+0.44}_{-0.31}\times10^{25}$ yr
in \cite{KlapdorKleingrothaus:2004wj}. The study of vacuum stability gives $M_D \le 24.36$ GeV. Using the
values for $ T_{1/2}$ and $M_D$, we can put the limit on the mass of the heavy neutrino,
\bea
M_R \leq 4.5~ \rm{TeV}
\eea
\subsection{Lepton flavor violation}
\begin{figure}[ht]
\begin{center}
\includegraphics[width=0.4\textwidth]{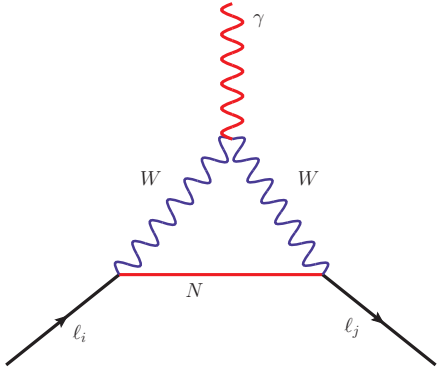}
\end{center}
\caption{\textit{Lepton Flavour Violating process $\ell_i \to \ell_j \gamma$.}}
\label{lfv1}
\end{figure}

The mixing of active neutrinos with heavy neutrinos can give rise to lepton flavour violations (LFV) like $\mu\rightarrow e \gamma$
 as shown in Fig.~\ref{lfv1}, if we generalise $\mathcal{M}_R$ matrix to contain off-diagonal terms. Assuming
the structure of $\mathcal{M}_R$ matrix as:
\begin{eqnarray}
\mathcal{M}_{R}^{-1}=M_R^{-1}\left(\begin{array}{ccc}
1 & \epsilon_1 & \epsilon_2 \\
\epsilon_1 & 1 & \epsilon_3 \\
\epsilon_2 & \epsilon_3 & 1
\end{array} \right),
\end{eqnarray}
where $\epsilon_i$s can be chosen to satisfy the correct light neutrino maxing angles.

We know among the $\ell_i \to \ell_j \gamma$ type LFV decays the $\mu \to e \gamma$ holds the most stringent bound on its decay branching ratio (BR)
which is $2.4 \times 10^{-12}$ (Present) \cite{mu2eg1}, and  $1.0 \times 10^{-13}$ (Future) \cite{mu2eg2}.

We estimate the branching ratio of this
process from vacuum stability and check its compatibility with the existing direct bounds.
This branching ratio for $\mu \to e \gamma$ is accompanied by the mixing $V_{li}\,(l= e,\mu)$ between
light to heavy neutrino \cite{lfv_references,Atre:2009rg}:
\begin{eqnarray}
{\rm Br}\left(\mu \,\, \rightarrow \,\,e \,\,\gamma\right)
=\frac{3\alpha}{8\pi}\left|\sum_i V_{e i} V_{\mu i}^{*}\, \hat{g}(r)\right|^2,
\end{eqnarray}
where $\hat{g}(r)=r\left(1-6r+3r^2+2r^3-6r^2\ln(r)\right)/(2\left(1-r\right)^4)$, and  $r=M_R^2/M_W^2$.

Again taking the constraint from vacuum stability $M_D \simeq 24.36$ GeV
\begin{eqnarray}
{\rm Br}\left(\mu \rightarrow e \gamma\right)= 2.82\times 10^{-10}\left(\frac{M_D}{ 24.36 ~{\rm GeV}}\right)^4\left(\frac{\rm TeV}{M_R}\right)^4.
\end{eqnarray}

Taking the experimental bound \cite{mu2eg1}  ${\rm Br}\left(\mu \rightarrow e \gamma\right)< 2.4 \times 10^{-12}$ and if  
$M_D \simeq 24.36$ GeV (in order to give a sizable contribution to ($0\nu \beta \beta$) and SSD signal at LHC) then 
we see that $M_R \geq 3.3$ TeV. This implies that in order to observe
a $(0 \nu  \beta \beta)$ or like-sign-dilepton signals where we need $M_R$ to be small, 
the texture of $\mathcal{Y}_\nu$ and $\mathcal{M}_R$ should be such that the $e-\mu$ flavour mixing is small.

\section{Same-Sign-Dilepton signal at LHC}
The processes for same-sign-dilepton (SSD)  production are similar to the
 neutrinoless double beta decay, see Fig.~\ref{0n2b}. These processes are of phenomenological importance as it 
involves both $e$ and $\mu$ . The signal is identified as the same-sign-dileptons + $N$ jets, $N>2$.
 The interaction vertices of the  heavy neutrino ($N_R$) are suppressed by the mixing parameters $\sim \mathcal{O}(Y_\nu v/M_R)$.
Assuming again a flavour diagonal $\mathcal{Y}_\nu$ and degenerate  $N_R$ we estimate the cross section for SSD at the LHC.

We have implemented this SM $\oplus$ Heavy Singlet neutrino model at Calchep \cite{calchep}, and estimated the cross-section 
for the process $p p\rightarrow e^{\pm} e^{\pm}+{\rm jets}$ and $p p\rightarrow \mu^{\pm} \mu^{\pm}+{\rm jets}$.
We have considered the range of $M_R$ to be 0.1-1 TeV, and no flavour structure for the simplification of study.
 It has been noted that in the Fig.~\ref{0n2b} (left)
the amplitude is suppressed more ($(M_D/M_R)^4$) than the other diagram Fig.~\ref{0n2b} (right) (here the suppression is
$\mathcal{O}(M_D/M_R)^2$. The choice of our $M_R$ is such that the mixing is much smaller than 1, and
that dictates us to work safely with the Fig.~\ref{0n2b} (right).

\begin{figure}[htb]
\begin{center}
\includegraphics[width=0.8\textwidth]{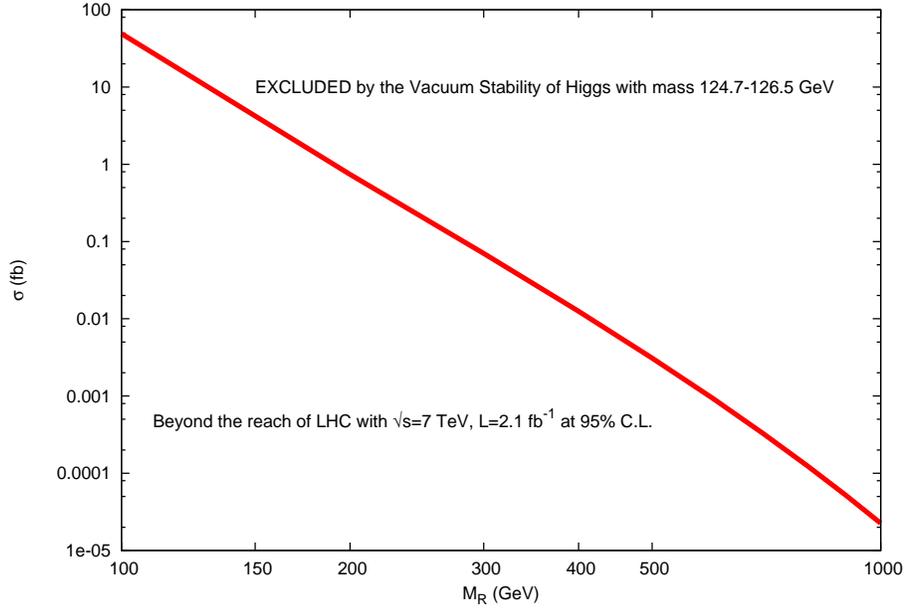}
\end{center}
\caption{\textit{Production cross-section Fig.~4 (right) with $\sqrt{s}$=7 \rm{TeV} in the LHC.}}
\label{cs7tev}
\end{figure}

In earlier section we have noted the maximum $M_D=Y_\nu v$ from the vacuum stability of the Standard Model Higgs field.
In this section we have use that limit and estimate the largest possible maximum cross-section for the process
Fig.~\ref{0n2b} (right) with two different sets of center of mass energy at the LHC. These two cross-sections
are calculated with center of mass energy ($\sqrt{s}$) 
7 TeV see Fig.~\ref{cs7tev} and 14 TeV see Fig.~7. 

\begin{figure}[htb]
\begin{center}
\includegraphics[width=0.8\textwidth]{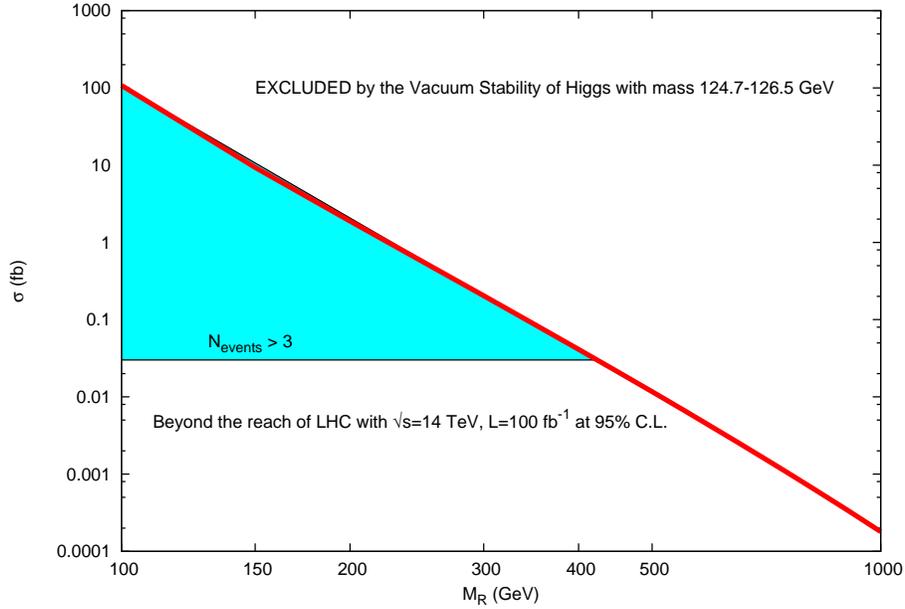}
\caption{\textit{Production cross-section of the process Fig.~4 (right) with $\sqrt{s}$=14 \rm{TeV} in the LHC.}}
\end{center}
\label{cs14tev_final}
\end{figure}

In recent paper by ATLAS \cite{atlas-2012} the Standard Model background has been estimated at 2.1 $fb^{-1}$ luminosity.
As shown in Fig.~\ref{cs14tev_final}, and Fig.~\ref{cs7tev} the vacuum stability puts a stringent bound on the production cross-section, 
through the $M_D$, the maximum allowed cross-section
is 49.02 fb at 7  TeV center of mass energy. This is the maximum cross-section that one
attains using no cuts. But due to the stringent constraint from the demand of vacuum stability the allowed cross-section is quite
small that yet LHC does not have enough data to see the process compared to the SM background \cite{atlas-2012}. 
Thus we have to wait for future data with 14 TeV center of mass energy and large integrated luminosity (L=$\int \mathcal{L} dt$ = $\sim$ 100 $fb^{-1}$).
The cross section for the SSD process at the LHC with $\sqrt{s}$=14 TeV is shown in  Fig.~\ref{cs14tev_final} 
and the region above the `thick (red)' line is disallowed by the vacuum stability.
In Fig.~\ref{cs14tev_final} we see that `shaded (cyan)' region is the one accessible  at the LHC with $\sqrt{s}$=14 TeV 
at L=100 $fb^{-1}$ considering atleast 3 events over the zero background, i.e, at 95\% C.L. Taking into account the vacuum stability condition it may 
be possible to observe SSD signal at LHC if $M_R <$ 400 GeV.

\section{Conclusion}
In this paper we have focused on the vacuum stability of the Higgs field in a specific scenario where the Standard Model
is extended by singlet Majorana fermions. We have studied the impact of such new field that couples to the light neutrinos via the
SM Higgs doublet on the RG evolution of the Higgs quartic coupling ($\lambda_h$), and we show that expectedly this new coupling ($Y_\nu$)
lowers the scale $\mu$ at which $\lambda_h(\mu)$ becomes negative. In this study the aim  is to find the maximum value of $Y_\nu$
which is compatible with the vacuum stability with heavy neutrino field having mass $M_R \sim$ TeV.
 
 We showed that the vacuum stability condition constrains the Dirac mass (which we have taken to be degenerate) to be 
 $M_D \leq 24.36$ GeV. We studied $\Delta L=2$ processes like $(0 \nu \beta \beta)$ and same-sign-dileptons at LHC and 
 lepton flavour violating processes like $\mu \rightarrow e \gamma$ taking into account the vacuum stability bound on 
$M_D$ which restrict the mixing between the light and heavy neutrinos which goes as $M_D/M_R$.
 
 We find that in order to observe $(0 \nu \beta \beta)$ signal which saturates the experimental bound 
$T_{1/2}= 2.23^{+0.44}_{-0.31}\times10^{25}$ yr \cite{KlapdorKleingrothaus:2004wj} the heavy neutrinos must have a mass $M_R < 4.5$ TeV.
 
 For the LFV process $\mu\rightarrow e \gamma$ if we assume $M_D$ at the largest possible value 24.36 GeV 
from vacuum stability (to maximise the chances for other signals), then we get the constraint $M_R > 3.3$ TeV. 
It may be possible to evade this bound on $M_R$ by choosing the texture of ${\mathcal M_D}$ and ${\mathcal M_R}$ matrices such that $e -\mu$ mixing is suppressed.
 
 Finally  estimated the maximal cross-section for the
signal, like same-sign-dilepton associated with jets imposing the vacuum stability condition.
We show that the data attained with  2.1 $fb^{-1}$ integrated luminosity cannot rule out right-handed neutrinos as 
the vacuum stability criterion shows that the dilepton signal would be way below the SM background.
It may be possible to observe the SSD at the LHC with  $\sqrt{s}$=14 TeV and integrated luminosity of 100 $fb^{-1}$
as long as $M_R< 400$GeV. If a larger signal is seen at the LHC then it would be a sign of new physics beyond SM + sterile right-handed neutrinos.

\section{Acknowledgements}
SM would like to thank Anjan Joshipura for useful suggestions and  discussions.
JC would like to thank Srubabati Goswami and Partha Konar for some useful discussions.



\begin{thebibliography}{52}

\bibitem{Atlas-cms} J. Incandela, CMS talk at Latest update in the search for the Higgs boson at CERN, July 4,
2012.; F. Gianotti, ATLAS talk at Latest update in the search for the Higgs boson at
CERN, July 4, 2012.


\bibitem{ATLAS:2012ae}
  G.~Aad {\it et al.} ATLAS Collaboration,
  Phys.\ Lett.\ B {\bf 710}, 49 (2012)
  arXiv:1202.1408 [hep-ex].

\bibitem{Chatrchyan:2012tx}
  S.~Chatrchyan {\it et al.}  CMS Collaboration,
  Phys.\ Lett.\ B {\bf 710}, 26 (2012)
  arXiv:1202.1488 [hep-ex].

\bibitem{Altarelli:1994rb}
  G.~Altarelli and G.~Isidori,
  Phys.\ Lett.\ B {\bf 337}, 141 (1994).

\bibitem{Holthausen:2011aa}
  M.~Holthausen, K.~S.~Lim and M.~Lindner,
  JHEP {\bf 1202}, 037 (2012)
  [arXiv:1112.2415 [hep-ph]].


\bibitem{EliasMiro:2011aa}
  J.~Elias-Miro, J.~R.~Espinosa, G.~F.~Giudice, G.~Isidori, A.~Riotto and A.~Strumia,
  Phys.\ Lett.\ B {\bf 709}, 222 (2012)
  [arXiv:1112.3022 [hep-ph]].

\bibitem{Xing:2011aa}
  Z.~z.~Xing, H.~Zhang and S.~Zhou,
  arXiv:1112.3112 [hep-ph].
\bibitem{Chetyrkin:2012rz}
  K.~G.~Chetyrkin and M.~F.~Zoller,
  JHEP {\bf 1206}, 033 (2012)
  [arXiv:1205.2892 [hep-ph]].

\bibitem{Bezrukov:2012sa}
  F.~Bezrukov, M.~Y.~.Kalmykov, B.~A.~Kniehl and M.~Shaposhnikov,
  arXiv:1205.2893 [hep-ph].

\bibitem{Degrassi:2012ry}
   G.~Degrassi, S.~Di Vita, J.~Elias-Miro, J.~R.~Espinosa, G.~F.~Giudice, G.~Isidori and A.~Strumia,
  arXiv:1205.6497 [hep-ph].

\bibitem{Casas:1999cd}
  J.~A.~Casas, V.~Di Clemente, A.~Ibarra and M.~Quiros,
  Phys.\ Rev.\ D {\bf 62}, 053005 (2000)
  [hep-ph/9904295].

\bibitem{Gogoladze:2008gf}
  I.~Gogoladze, N.~Okada and Q.~Shafi,
  Phys.\ Rev.\ D {\bf 78}, 085005 (2008)
  [arXiv:0802.3257 [hep-ph]].



\bibitem{Gogoladze:2008ak}
  I.~Gogoladze, N.~Okada and Q.~Shafi,
  Phys.\ Lett.\ B {\bf 668}, 121 (2008)
  [arXiv:0805.2129 [hep-ph]].

\bibitem{Chen:2012fa}
  C.~-S.~Chen and Y.~Tang,
  JHEP {\bf 1204}, 019 (2012)
  [arXiv:1202.5717 [hep-ph]].




\bibitem{Pilaftsis:1991ug}
  A.~Pilaftsis,
  Z.\ Phys.\ C {\bf 55}, 275 (1992)
  [hep-ph/9901206].


\bibitem{Rodejohann:2012px}
  W.~Rodejohann and H.~Zhang,
  arXiv:1203.3825 [hep-ph].

\bibitem{Type-1}
P. Minkowski, Phys. Lett. B67, 421 (1977); T. Yanagida, in Proc. Workshop on the baryon
number of the Universe and unifed theories, edited by O. Sawada and A. Sugamoto (1979),
p. 95; R. N. Mohapatra and G. Senjanovi´c, Phys. Rev. Lett. 44, 912 (1980); M. Gell-Mann,
P. Ramond, and R. Slansky, in Supergravity, edited by P. van Nieuwenhuizen and D. Freedman
(1979), p. 315.heavy


\bibitem{Grimus:2000vj}
  W.~Grimus and L.~Lavoura,
  JHEP {\bf 0011}, 042 (2000)
  [hep-ph/0008179].

\bibitem{Kersten:2007vk}
  J.~Kersten and A.~Y.~.Smirnov,
  Phys.\ Rev.\ D {\bf 76}, 073005 (2007)
  [arXiv:0705.3221 [hep-ph]].
\bibitem{Hettmansperger:2011bt}
  H.~Hettmansperger, M.~Lindner and W.~Rodejohann,
  JHEP {\bf 1104}, 123 (2011)
  [arXiv:1102.3432 [hep-ph]].

\bibitem{Atre:2009rg}
  A.~Datta, M.~Guchait and A.~Pilaftsis,
  Phys.\ Rev.\ D {\bf 50}, 3195 (1994)
  [hep-ph/9311257];

  F.~M.~L.~Almeida, Jr., Y.~D.~A.~Coutinho, J.~A.~Martins Simoes and M.~A.~B.~do Vale,
  Phys.\ Rev.\ D {\bf 62}, 075004 (2000)
  [hep-ph/0002024];

  O.~Panella, M.~Cannoni, C.~Carimalo and Y.~N.~Srivastava,
  Phys.\ Rev.\ D {\bf 65}, 035005 (2002)
  [hep-ph/0107308];

  T.~Han and B.~Zhang,
  Phys.\ Rev.\ Lett.\  {\bf 97}, 171804 (2006)
  [hep-ph/0604064];

  A.~Atre, T.~Han, S.~Pascoli and B.~Zhang,
  JHEP {\bf 0905}, 030 (2009)
  [arXiv:0901.3589 [hep-ph]].

\bibitem{Ibarra:2010xw}
  A.~Ibarra, E.~Molinaro and S.~T.~Petcov,
  JHEP {\bf 1009}, 108 (2010)
  [arXiv:1007.2378 [hep-ph]].
  
\bibitem{Jones}  
 C.~Ford, I.~Jack and D.~R.~T.~Jones,
  Nucl.\ Phys.\ B {\bf 387}, 373 (1992)
  [Erratum-ibid.\ B {\bf 504}, 551 (1997)]
  [hep-ph/0111190].



\bibitem{Hambye:1996wb}
  T.~Hambye and K.~Riesselmann,
  Phys.\ Rev.\ D {\bf 55}, 7255 (1997)
  [hep-ph/9610272].



\bibitem{topmass-measurement}
  Tevatron Electroweak Working Group and CDF and D0 Collaborations,
  arXiv:1107.5255 [hep-ex].

\bibitem{topmass-higgsmass}
S.~Alekhin, A.~Djouadi and S.~Moch,
  arXiv:1207.0980 [hep-ph].

\bibitem{Mitra:2011qr}
  M.~Mitra, G.~Senjanovic and F.~Vissani,
  Nucl.\ Phys.\ B {\bf 856}, 26 (2012)
  [arXiv:1108.0004 [hep-ph]].


\bibitem{KlapdorKleingrothaus:2004wj}
  H.~V.~Klapdor-Kleingrothaus, I.~V.~Krivosheina, A.~Dietz and O.~Chkvorets,
  Phys.\ Lett.\ B {\bf 586}, 198 (2004)
  [hep-ph/0404088].


\bibitem{mu2eg1}
  J.~Adam {\it et al.}  [MEG Collaboration],
  Phys.\ Rev.\ Lett.\  {\bf 107} (2011) 171801
  [arXiv:1107.5547 [hep-ex]].

\bibitem{mu2eg2}
  A.~Hoecker,
  arXiv:1201.5093 [hep-ph].


\bibitem{lfv_references} 
  E.~Ma and A.~Pramudita,
  Phys.\ Rev.\ D {\bf 24}, 1410 (1981);
P.~Langacker and D.~London,
  Phys.\ Rev.\ D {\bf 38}, 907 (1988);
D.~Tommasini, G.~Barenboim, J.~Bernabeu and C.~Jarlskog,
  Nucl.\ Phys.\ B {\bf 444}, 451 (1995)
  [hep-ph/9503228].



\bibitem{calchep}
  A.~Pukhov,
  arXiv:hep-ph/0412191.



\bibitem{atlas-2012}
G.~Aad {\it et al.}  [ATLAS Collaboration],
  arXiv:1203.5420 [hep-ex].





\end{thebibliography}
\end{document}